\documentclass[10pt]{amsart}
\usepackage{amsmath}
\usepackage{amssymb}
\usepackage{amsfonts}
\usepackage{euscript}
\usepackage{amsthm,amsxtra}
\usepackage{amsfonts}
\usepackage{nicefrac}
\theoremstyle{plain}
\newtheorem{theorem}{Theorem}[section]

\newtheorem{proposition}{Proposition}[section]
\theoremstyle{definition}

\theoremstyle{remark}

\newcommand{\PV}{${\rm P}_{\rm V}\;$}

\newcommand{\PIIIa}{${\rm P}_{\rm III^{\prime}}\;$}

\newcommand{\PVI}{${\rm P}_{\rm VI}\;$}

\newcommand{\ZZ}{\mathbb Z}

\newcommand{\CC}{\mathbb C}
\newcommand{\TT}{\mathbb T}

\newcommand{\half}{\nicefrac{1}{2}\:}

\newcommand{\quarter}{\nicefrac{1}{4}\:}
\newcommand{\tquarter}{\nicefrac{3}{4}}

\begin{document}
%\begin{spacing}{1.5}

\vspace{4cm}
\noindent
{\bf Discrete Painlev\'e equations, Orthogonal Polynomials on the Unit Circle and 
$N$-recurrences for averages over $U(N)$ -- \PIIIa and \PV $\tau$-functions}

\vspace{5mm}
\noindent
P.J.~Forrester and N.S.~Witte${}^\dagger$

\noindent
Department of Mathematics and Statistics
${}^\dagger$(and School of Physics), \\
University of Melbourne, Victoria 3010, Australia \\
email: p.forrester@ms.unimelb.edu.au; n.witte@ms.unimelb.edu.au

\small
\begin{quote}
In this work we show that the $ N\times N $ Toeplitz determinants with the symbols 
$ z^{\mu}\exp(-\frac{1}{2}\sqrt{t}(z+1/z)) $ and 
$ (1+z)^{\mu}(1+1/z)^{\nu}\exp(tz) $ -- known $\tau$-functions for the \PIIIa and 
\PV systems -- 
are characterised by nonlinear recurrences for the reflection coefficients of the
corresponding orthogonal polynomial system on the unit circle. It is shown that
these recurrences are entirely equivalent to the discrete Painlev\'e equations
associated with the degenerations of the rational surfaces 
$ D^{(1)}_{6} \to E^{(1)}_{7} $ (discrete Painlev\'e {\rm II}) and 
$ D^{(1)}_{5} \to E^{(1)}_{6} $ (discrete Painlev\'e {\rm IV}) respectively
through the algebraic methodology based upon of the affine Weyl group symmetry of the 
Painlev\'e system, originally due to Okamoto. In addition it is shown that the 
difference equations derived by methods based upon the Toeplitz lattice and Virasoro
constraints, when reduced in order by exact summation, are equivalent to our
recurrences. 
Expressions in terms of generalised hypergeometric functions
$ {{}^{\vphantom{(1)}}_0}F^{(1)}_1, {{}^{\vphantom{(1)}}_1}F^{(1)}_1 $ are given for
the reflection coefficients respectively.  
\end{quote}

\section{Introduction}
\setcounter{equation}{0}

There are now at least three approaches to systematically obtain recurrences for
random matrix averages corresponding to $\tau$-functions for Painlev\'e systems.
One, exploited by the present authors \cite{FW_2003a}, is to use the theory of
Schlesinger transformations within the $\tau$-function theory of Painlev\'e systems.
Another, due to Borodin \cite{Bo_2001}, is based on a discrete analogue of the 
Riemann-Hilbert problem \cite{B_2000},\cite{BB_2002}. A third due to Adler and 
van Moerbeke \cite{AvM_2002} is based on the theory of the integrable Toeplitz 
lattice and Virasoro constraints. In the work of the present authors, 
and of Borodin, there is an explicit connection with the discrete Painlev\'e
equations. Thus in all cases the recurrences for the $\tau$-functions involve auxiliary
quantities which satisfy discrete Painlev\'e equations. However the recurrences 
obtained in the work of Adler and van Moerbeke were not, in general, related to 
discrete Painlev\'e equations. This then immediately raises the question as to the 
relationship between the recurrences obtained by Adler and van Moerbeke, and the 
discrete Painlev\'e recurrences. In this work, for the recurrences relating to
$\tau$-functions for the Painlev\'e $\rm III'$ and the Painlev\'e V systems,
we will answer this question by showing that in fact the recurrences obtained in
\cite{AvM_2002} are transformed versions of the discrete Painlev\'e equations. 
Moreover, we will show that the recurrences of Adler and van Moerbeke, obtained 
from their theory of the integrable Toeplitz lattice, follow from an approach 
based on the theory of orthogonal polynomials on the unit circle with 
semi-classical weights. In the situation of a general weight the recurrence relations
for the various coefficients appearing in the orthogonal polynomial system are known 
as Freud or Laguerre-Freud equations \cite{Fr_1976,La_1972b}. This theory then provides 
a fourth approach to systematically obtain recurrences for random matrix averages 
corresponding to $\tau$-functions for Painlev\'e systems.

In this work attention will be focused on an orthogonal polynomial approach to 
$\tau$-functions for the Painlev\'e $\rm III'$ and Painlev\'e V systems defined 
as averages over the eigenvalue probability density function for the unitary group
$U(N)$ with Haar (uniform) measure (see e.g.~\cite[Chapter 2]{rmt_Fo}),   
\begin{equation}\label{1.3}
   {1 \over (2 \pi )^N N!} \prod_{1 \le j < k \le N} | z_k - z_j |^2, \qquad
   (z_j := e^{i \theta_j}, -\pi < \theta_j \le \pi, j=1,\ldots,N) 
\end{equation} 
These $\tau$-functions are 
\begin{align}
  \tau^{\rm III'}[N](t;\mu) 
  & := t^{-N\mu/2} \Big\langle \prod_{l=1}^N z_l^\mu
       e^{{1 \over 2}\sqrt{t}(z_l+z_l^{-1})} \Big\rangle_{U(N)}
  \label{III_Toep} \\
  \tau^{\rm V}[N](t;\mu,\nu)  
  & := \Big\langle \prod_{l=1}^N (1+z_l)^\mu (1+1/z_l)^{\nu}e^{t z_l} \Big\rangle_{U(N)}
  \label{V_Toep}
\end{align}
where in (\ref{III_Toep}) it is assumed $ \mu \in \ZZ $ while in (\ref{V_Toep}) it is
assumed $ \mu, \nu \in \ZZ_{\geq 0} $. By noting the identity
\begin{equation}
   (1+z)^{\mu}(1+1/z)^{\nu} = z^{(\mu-\nu)/2}|1+z|^{\mu+\nu}
\label{V_modulus}
\end{equation}
we can rewrite (\ref{V_Toep}) to read 
\begin{equation}
  \tau^{\rm V}[N](t;\mu,\nu) = 
  \Big\langle \prod_{l=1}^N z_l^{(\mu-\nu)/2}|1+z_l|^{\mu+\nu}e^{t z_l} \Big\rangle_{U(N)}
  \label{V_Toep2} 
\end{equation}
which is well defined for $ \Re(\mu+\nu) > -1 $. Associated with (\ref{III_Toep}) and 
(\ref{V_Toep2}) are the weight functions
\begin{equation}
  z^\mu e^{{1 \over 2}\sqrt{t}(z+z^{-1})},
  z^{(\mu-\nu)/2}|1+z|^{\mu+\nu}e^{tz},    
\end{equation}
on the unit circle $ z \in \mathbb{T} $ which have the special property of being 
semi-classical. On this latter point, analogous to the use of the term classical weight
function for orthogonal polynomials on the line (see e.g.~\cite{AFNV_2000}) 
we will call a weight function $w(z)$ on the unit circle classical if
its logarithmic derivative is of the form $g(z)/f(z)$ with $g(z)$
a polynomial of degree $\le 1$, and $f(z)$ is a polynomial of degree $\le 2$. This
gives $ w(z) = z^{(\mu-\nu)/2}|1+z|^{\mu+\nu} $ as the only classical weight on the 
unit circle. The weights closest to classical with respect to the degree of the 
corresponding polynomials $g$ and $f$ are 
\begin{equation}
  z^{\mu}e^{{1 \over 2}\sqrt{t}(z+z^{-1})}, \qquad
  z^{(\mu-\nu)/2}|1+z|^{\mu+\nu}e^{tz},
  \label{SC_wgt}
\end{equation}
and are to be termed semi-classical. In the case of the first weight in (\ref{SC_wgt})
with $ \mu=0 $, it is known that the corresponding orthogonal polynomials satisfy
special recurrence relations \cite{IW_2001,Ma_2000} which lead to a
recurrence for the corresponding $ U(N) $ average. We will show that
this is also true of the general form of the first weight in (\ref{SC_wgt}), as well
as the second weight in (\ref{SC_wgt}), thus giving recurrences for 
$ \tau^{\rm III'}[N] $ and $ \tau^{\rm V}[N] $. In general these are different 
from those obtained in the Painlev\'e systems approach, but rather coincide with 
recurrences obtained by Adler and van Moerbeke from their theory of the
Toeplitz lattice and its Virasoro algebra \cite{AvM_2002}. As already remarked,
we are able to show that after appropriate transformations, the recurrences of the 
two approaches coincide.

Another theme we wish to develop is the solution of the recurrences associated 
with (\ref{III_Toep}) and (\ref{V_Toep2}) in terms of generalised hypergeometric 
functions based on Schur polynomials. To define the latter let 
$ \kappa = (\kappa_1,\kappa_2, \ldots, \kappa_N) $ denote a partition so that the
$ \kappa_i $'s are non-negative integers with 
$ \kappa_1 \geq \kappa_2 \geq \cdots \geq \kappa_N \geq 0 $,
let $ s_{\kappa}(t_1,\ldots,t_N) $ denote the Schur symmetric polynomial, and define
the generalised Pochhammer symbol
\begin{equation}
   [a]^{(1)}_{\kappa} := \prod^{N}_{j=1}(a-j+1)_{\kappa_j}, \qquad
   (a)_l := a(a+1) \ldots (a+l-1).
\end{equation}
Also, with $ (i,j)\in \kappa $ referring to a node in the Young diagram of $ \kappa $
and $ a(i,j), l(i,j) $ the corresponding arm and leg lengths respectively 
(see \cite{Ma_1995}, define the hook length 
\begin{equation}
   h_{\kappa} = \prod_{(i,j) \in \kappa} [a(i,j)+l(i,j)+1] .
\end{equation}
With this notation the generalised hypergeometric series of interest is defined 
through a series representation \cite{Ya_1992,Ka_1993}
\begin{equation}
  {}^{\vphantom{(1)}}_{p}F^{(1)}_{q}(a_1,\ldots,a_p;b_1,\ldots,b_q;t_1,\ldots,t_N)
  = \sum^{\infty}_{\kappa} 
    {[a_1]^{(1)}_{\kappa}\cdots [a_p]^{(1)}_{\kappa} \over 
     [b_1]^{(1)}_{\kappa}\cdots [b_q]^{(1)}_{\kappa}}
    {s_{\kappa}(t_1,\ldots,t_N) \over h_{\kappa}}
  \label{GenHyp}
\end{equation}
for $ p,q \in \ZZ_{\geq 0} $. 
The superscript $ (1) $ in $ {}^{\vphantom{(1)}}_{p}F^{(1)}_{q} $ indicates 
(\ref{GenHyp}) is a special case of a hypergeometric series based on Jack polynomials
and depends on a complex parameter $ d $, the case $ d=1 $ corresponding to
(\ref{GenHyp}).

In Section 2 we present some general formulas from the Szeg\"o theory of orthogonal 
polynomials on the unit circle \cite{ops_Sz} required for our study of (\ref{III_Toep}) 
and (\ref{V_Toep2}), the later being carried out in Sections 3 and 4 respectively.
In studying (\ref{III_Toep}) and (\ref{V_Toep2}) we first present the recurrence
scheme following from the work of Adler and van Moerbeke \cite{AvM_2002}, and then 
proceed to show how the same schemes, and ones of lower order, can be derived
from the theory of orthogonal polynomials on the unit circle. We make note of the 
solution of these recurrences in terms of the generalised hypergeometric functions,
before revising the recurrence schemes following from the Painlev\'e systems 
approach. The latter do not coincide with the recurrences resulting from the 
orthogonal polynomial approach, however we show that after an appropriate
transformation of variables the recurrences are in fact equivalent.
We label the particular discrete equations that arise according to the 
unambiguous algebraic-geometric classification of Sakai \cite{Sa_2001} by
association with a degeneration of a particular rational surface into 
another, rather than the previous names employed.

\section{Orthogonal Polynomials on the Unit Circle}
\setcounter{equation}{0}

We will consider orthogonal polynomials with respect to a complex weight function 
$ w(z) $, analytic in the cut complex $ z $-plane. The latter specification means
$ w(z) $ possesses a Fourier expansion
\begin{equation}
  w(z) = \sum_{k=-\infty}^{\infty} w_{k}z^k, \quad
  w_{k} = \int_{\TT} {dz \over 2\pi iz} w(z)z^{-k},
\label{ops_Fourier}
\end{equation}
where $ \TT $ denotes the unit circle $ |z|=1 $, appropriately deformed so not
to cross the cut, and $ z=e^{i\theta}, \theta \in (-\pi,\pi] $. 
For $ \epsilon = 0,\pm 1 $ we define the Toeplitz determinants
\begin{equation}
   I^{\epsilon}_{N}[w] 
  := \det \left[ \int_{\TT} {dz \over 2\pi iz} w(z)z^{\epsilon-j+k}
           \right]_{0 \leq j<k \leq N-1}
  = \det \left[ w_{-\epsilon+j-k} \right]_{0 \leq j<k \leq N-1}.
\label{ops_Idefn}
\end{equation}
In the case $ \epsilon = 0 $, and $ \TT $ the unit circle without deformation,
by virtue of the identity
\begin{equation}
  \det \left[ \int_{\TT} {dz \over 2\pi iz} w(z)z^{\epsilon+j-k}
           \right]_{0 \leq j<k \leq N-1} =
  \Big\langle \prod_{l=1}^N z^{\epsilon}_lw(z_l) \Big\rangle_{U(N)}
\label{ops_Uint}
\end{equation}
we see that (\ref{III_Toep}) and (\ref{V_Toep2}), in the cases $ \mu \in \ZZ $ and
$ \mu, \nu \in \ZZ_{\geq 0} $ at least, can be expressed as
$ I^0_N[w] $ with $ w(z) $ as in (\ref{SC_wgt}). In certain circumstances the 
weight is real and positive $ \overline{w(z)} = w(z) $ and thus the Toeplitz matrix 
is Hermitian, $ \bar{w}_k = w_{-k} $, but in general this will not be true.

If $ I^{0}_{N}[w] $ is non-zero for each $ N=1,2, \ldots $ then there exists 
a system of orthogonal polynomials 
$ \{ \phi_n(z), n = 0,1, \ldots \} $ with the orthonormality property 
\begin{equation}
  \int_{\TT} {dz \over 2\pi iz}\tilde{w}(z)\phi_m(z)\overline{\phi_n(z)}  
   = \delta_{m,n}
\label{ops_onorm}
\end{equation}
where 
\begin{equation}
   \tilde{w}(z) := {w(z) \over w_0} .
\label{ops_normal}
\end{equation}
We introduce special notation for the various coefficients in $ \phi_n(z) $
according to
\begin{equation}
   \phi_n(z) = \kappa_n z^n + l_n z^{n-1}+ m_n z^{n-2} + \ldots + \phi_n(0)
             = \sum^{n}_{j=0} c_{n,j}z^j,
\label{ops_coeff}
\end{equation}
where without loss of generality $ \kappa_n $ is chosen to be real and positive.
We also define the reciprocal polynomial by
\begin{equation}
   \phi^{*}_n(z) := z^n\bar{\phi}(1/z) = \sum^{n}_{j=0} \bar{c}_{n,j}z^{n-j},
\end{equation}
where $ \bar{c} $ denotes the complex conjugate.
A fundamental quantity is the ratio $ r_n = \phi_n(0)/\kappa_n $,
known as reflection coefficients because of their role in the scattering theory
formulation of orthogonal polynomial systems on the unit circle. 

From the Szeg\"o theory \cite{ops_Sz} these coefficients and their complex 
conjugates are related to the above Toeplitz determinants (\ref{ops_Uint}) by
\begin{equation}
  r_N = (-1)^N{ I^{1}_{N}[w] \over I^{0}_{N}[w]}, \quad
  \bar{r}_N = (-1)^N{ I^{-1}_{N}[w] \over I^{0}_{N}[w]}.
\label{ops_Refl}
\end{equation}
In the case that $ w(z) $ is not real, $ \bar{r}_N $ (notwithstanding the 
notation) is not the complex conjugate of $ r_N $ but rather an independent 
variable. Note that the normalisation (\ref{ops_normal}) implies that 
$ \kappa_0 = 1 $ and thus $ r_0 = \bar{r}_0 = 1 $.
Knowledge of $ \{ r_N \}_{N=0,1,\ldots} $, $ \{ \bar{r}_N \}_{N=0,1,\ldots} $
is sufficient to compute $ \{ I^{0}_{N}[w] \}_{N=0,1,\ldots} $. For this one
uses the general formula \cite{ops_Sz}
\begin{equation}
   {I^{0}_{N+1}[w] I^{0}_{N-1}[w] \over (I^{0}_{N}[w])^2}
   = 1 - r_{N}\bar{r}_N.
\label{ops_I0}
\end{equation}
Our fundamental task is then to obtain recurrences determining the $ r_N $ and
$ \bar{r}_N $.

For this purpose we require further formulae from the Szeg\"o theory. First, 
as a consequence of the orthogonality condition we have the mixed linear 
recurrence relations for $ \phi_{n} $ and $ \phi^*_{n} $,
\begin{align}
  \kappa_n  \phi_{n+1}(z)
   & = \kappa_{n+1}z \phi_{n}(z)+\phi_{n+1}(0) \phi^*_n(z)
  \label{ops_rr:a} \\
  \kappa_n \phi^*_{n+1}(z)
   & = \kappa_{n+1} \phi^*_{n}(z)+\bar{\phi}_{n+1}(0) z\phi _n(z) ,
  \label{ops_rr:b}
\end{align}
as well as the three-term recurrences
\begin{align}
  \kappa_n\phi_n(0)\phi_{n+1}(z) + \kappa_{n-1}\phi_{n+1}(0)z\phi_{n-1}(z)
   & = (\kappa_{n}\phi_{n+1}(0)+\kappa_{n+1}\phi_{n}(0)z)\phi_n(z)
  \label{ops_ttr:a} \\
  \kappa_n\bar{\phi}_n(0)\phi^*_{n+1}(z) 
  + \kappa_{n-1}\bar{\phi}_{n+1}(0)z\phi^*_{n-1}(z)
   & = (\kappa_{n}\bar{\phi}_{n+1}(0)z+\kappa_{n+1}\bar{\phi}_{n}(0))\phi^*_n(z) .
  \label{ops_ttr:b} 
\end{align}
From the latter one can derive the analogue of the Christoffel-Darboux summation 
formula
\begin{align}
  \sum^{n}_{j=0} \phi_j(z)\overline{\phi_j(\zeta)} 
  & =
  { \phi^*_n(z)\overline{\phi^*_n(\zeta)}
   -z\bar{\zeta}\phi_n(z)\overline{\phi_n(\zeta)} \over 1-z\bar{\zeta} }
  \label{ops_CD:a} \\
  & = 
  { \phi^*_{n+1}(z)\overline{\phi^*_{n+1}(\zeta)}
   -\phi_{n+1}(z)\overline{\phi_{n+1}(\zeta)} \over 1-z\bar{\zeta} } ,
  \label{ops_CD:b}
\end{align}
for $ z\bar{\zeta} \not= 1 $.
Identities from the Szeg\"o theory that relate the leading coefficients back 
to the reflection coefficients are
\begin{align}
   \kappa_n^2 & = \kappa_{n-1}^2 + |\phi_n(0)|^2,
   \label{ops_kappa} \\
   {l_{n} \over \kappa_{n}} & = \sum^{n-1}_{j=0} r_{j+1}\bar{r}_{j},
   \label{ops_l} \\
   {m_{n} \over \kappa_{n}} 
    & = \sum^{n-1}_{j=0} r_{j+1}\bar{r}_{j-1} 
                       + r_{j+1}\bar{r}_{j}{l_{j-1} \over \kappa_{j-1}}.
   \label{ops_m}
\end{align}
Another relevant formula is 
\begin{equation}
    I^{0}_{N}[w] = \prod^{N-1}_{j=0} \kappa_j .
\label{ops_I0kappa}
\end{equation}
Finally, with $ \pi_{n} $ denoting an arbitrary polynomial in the linear space of 
polynomials with degree at most $ n $, we can check from the structure 
(\ref{ops_coeff}) that
\begin{align}
\begin{split}
  z\phi_n(z) & = {\kappa_{n} \over \kappa_{n+1}} \phi_{n+1}(z) 
       + \left({l_{n} \over \kappa_{n}}-{l_{n+1} \over \kappa_{n+1}}\right) \phi_{n}(z)
  \\
  & \quad
       + \left\{ {l_{n} \over \kappa_{n-1}}
                 \left( {l_{n+1} \over \kappa_{n+1}}-{l_{n} \over \kappa_{n}}\right)
                + {m_{n} \over \kappa_{n-1}}-{m_{n+1} \over \kappa_{n+1}}
                  {\kappa_{n} \over \kappa_{n-1}} \right\} \phi_{n-1}(z)
       + \pi_{n-2}
  \\
  z^2\phi_n(z) & ={\kappa_{n} \over \kappa_{n+2}} \phi_{n+2}(z) 
       + \left({l_{n} \over \kappa_{n+1}}-{l_{n+2} \over \kappa_{n+2}}
               {\kappa_{n} \over \kappa_{n+1}}\right) \phi_{n+1}(z)
  \\
  & \quad
       + \left\{ {l_{n+1} \over \kappa_{n+1}}
                 \left( {l_{n+2} \over \kappa_{n+2}}-{l_{n} \over \kappa_{n}}\right)
                + {m_{n} \over \kappa_{n}}-{m_{n+2} \over \kappa_{n+2}}
                  \right\} \phi_{n}(z) + \pi_{n-1} 
  \\
  \phi'_n(z) 
  & = n{\kappa_{n} \over \kappa_{n-1}} \phi_{n-1}(z) + \pi_{n-2} 
  \\
  z\phi'_n(z)
  & = n \phi_n(z) - {l_{n} \over \kappa_{n-1}} \phi_{n-1}(z)
                  + \pi_{n-2}
  \\
  z^2\phi'_n(z)
  & = n{\kappa_{n} \over \kappa_{n+1}} \phi_{n+1}(z)
         + \left\{(n-1){l_{n} \over \kappa_{n}}-n{l_{n+1} \over \kappa_{n+1}}
           \right\} \phi_n(z) + \pi_{n-1}
\end{split}
\label{ops_prod}
\end{align}
where $ ' $ denotes the derivative with respect to $ z $.

How we use the above formulae to produce recurrences for $ r_N $ and $ \bar{r}_N $
differs for the orthogonal polynomial systems corresponding to the two 
different weights in (\ref{SC_wgt}). Let us then treat the two weights 
separately.
Let us refer to the order of a difference equation in $ r_n, \bar{r}_n $ as 
$ q/p $ where $ q\in \mathbb{Z}_{\geq 0} $ refers to the order of $ r_n $ while 
$ p\in \mathbb{Z}_{\geq 0} $ refers to the order of $ \bar{r}_n $.
 
\section{The \PIIIa System}
\setcounter{equation}{0}

For non-integer values of $ \mu $ the weight function
\begin{equation}
   w(z) = z^{\mu} e^{\frac{1}{2}\sqrt{t}(z+z^{-1})} ,
\label{III_wgt}
\end{equation}
has a branch point at $ z=0 $. Cutting the complex plane along the negative real 
axis $ (-\infty,0] $ we deform the contour in (\ref{ops_Fourier}) to the contour
starting at $ -\infty $, running along the real axis on the negative imaginary
side to $ z=-1 $, following the circle $ |z|=1 $ in the anticlockwise direction 
to return to $ z=-1 $ on the positive imaginary side, then returning to 
$ -\infty $ along this side of the negative real axis. This contour is standard in
the theory of the Bessel function (see \cite{WW_1965},pg. 363).
Denoting this contour by $ \EuScript{C} $, and noting that the integral representation
of the Bessel function of pure imaginary argument gives
\begin{equation}
  \int_{\EuScript{C}}{dz \over 2\pi iz}w(z) = I_{\mu}(\sqrt{t}) ,
\label{Ibessel}
\end{equation}
we see that 
\begin{equation}
  I^{\epsilon}_N[w] = \det[ I_{\mu+\epsilon+j-k}(\sqrt{t}) ]_{j,k=1,\dots,N}.
\label{III_IToep}
\end{equation}
For general $ \mu $ we then define
\begin{equation}
  \tau^{\rm III'}[N](t;\mu) = \det[ I_{\mu+j-k}(\sqrt{t}) ]_{j,k=1,\dots,N} ,
\label{III_tau}
\end{equation}
which is consistent with (\ref{III_Toep}) in the case $ \mu \in \ZZ $.

Recurrences for the reflection coefficients $ r_N, \bar{r}_N $ in the case of the
weight (\ref{III_wgt}) can be deduced from the work of Adler and van Moerbeke 
\cite{AvM_2002}. In their case 3, one specialises their parameters 
$ d_1 = d_2 = \gamma_1' = \gamma_2' = \gamma_1'' = \gamma_2'' = 0 $, $ \gamma = \mu $ 
and one sets $ P_1(z) = P_2(z) = \half\sqrt{t}z $. Then making the identification 
$ r_{N} = x_{N} $, $ \bar{r}_{N} = y_{N} $ it follows from \cite{AvM_2002}, 
eq. (0.0.17) that
\begin{align}
   \half\sqrt{t}v_{N}(r_{N+1}+r_{N-1}) + Nr_{N} & = 0,
   \label{AvM_III:a} \\
   \half\sqrt{t}v_{N}(\bar{r}_{N+1}+\bar{r}_{N-1}) + N\bar{r}_{N} & = 0,
   \label{AvM_III:b}
\end{align}
where $ v_{N} := 1-r_{N}\bar{r}_{N} $. After specifying the initial conditions
\begin{equation}
   r_{0} = \bar{r}_{0} = 1, \quad
   r_{1} = -{I_{\mu+1}(\sqrt{t}) \over I_{\mu}(\sqrt{t})}, \quad
   \bar{r}_{1} = -{I_{\mu-1}(\sqrt{t}) \over I_{\mu}(\sqrt{t})} ,
\label{III_rInitial}
\end{equation}
which follow from (\ref{ops_Refl}) and (\ref{III_IToep}), the recurrences 
(\ref{AvM_III:a}) and (\ref{AvM_III:b}) uniquely determine $ r_N, \bar{r}_N $ for 
$ N=2,3,\ldots $. We note that the order of (\ref{AvM_III:a}) and (\ref{AvM_III:b})
is $ 2/0 $ and $ 0/2 $ respectively and the parameter $ \mu $ does not appear 
explicitly. In addition we observe that (\ref{AvM_III:a}) and (\ref{AvM_III:b})
have the familiar form of the discrete Painlev\'e equation associated with 
degeneration of the rational surfaces $ D^{(1)}_6 \to E^{(1)}_7 $ \cite{Sa_2001}
(discrete Painlev\'e {\rm II}). 
We now seek a derivation of (\ref{AvM_III:a}) and (\ref{AvM_III:b}) using 
the formulae of Section 2 specialised to the weight (\ref{III_wgt}). In addition
we will show that the orthogonal polynomial theory can be used to derive a pair of
coupled difference equations for $ r_N, \bar{r}_N $, both of order $ 1/1 $. It will 
then be shown how (\ref{AvM_III:a},\ref{AvM_III:b}) can be deduced from these 
equations.

\begin{proposition}\label{III_rProp}
The reflection coefficients (\ref{ops_Refl}) corresponding to the Toeplitz 
determinants (\ref{III_IToep}) satisfy the coupled $ 1/1 $ order recurrences
\begin{align}
   \half\sqrt{t}(r_{N+1}\bar{r}_{N}+r_{N}\bar{r}_{N-1}) 
  + N{r_{N}\bar{r}_{N} \over 1-r_{N}\bar{r}_{N}} - \mu & = 0,
  \label{III_rRecur:a} \\
   \half\sqrt{t}(\bar{r}_{N+1}r_{N}+\bar{r}_{N}r_{N-1}) 
  + N{r_{N}\bar{r}_{N} \over 1-r_{N}\bar{r}_{N}} + \mu & = 0,
  \label{III_rRecur:b}
\end{align}
with the initial conditions (\ref{III_rInitial}).
\end{proposition}
\begin{proof}
We adapt a method due to Freud \cite{Fr_1976}, and consider two different ways to 
evaluate the integral
\begin{equation}
  J_{1} := \int_{\EuScript{C}} {dz \over 2\pi iz} 
       z^2w'(z)\phi_{N}(z)\overline{\phi_{N+1}(z)} .  
\label{III_J1}
\end{equation}
Integrating this by parts, employing (\ref{ops_prod}) and the orthogonality 
conditions shows
\begin{equation}
 J_{1} = -(N+1){\kappa_{N} \over \kappa_{N+1}}
         +(N+1){\kappa_{N+1} \over \kappa_{N}}.
\label{III_J2}
\end{equation}
Alternatively we note from (\ref{III_wgt}) that
\begin{equation}                                         
  {w' \over w} = {\mu \over z} + \half\sqrt{t}(1-{1 \over z^2}),
\end{equation}
substituting this in (\ref{III_J1}), one can again use (\ref{ops_prod}) and the 
orthogonality conditions to show
\begin{equation}
  J_{1} = \mu{\kappa_{N} \over \kappa_{N+1}}
  + \half\sqrt{t}\left( {l_{N} \over \kappa_{N+1}} 
                      - {l_{N+2} \over \kappa_{N+2}}{\kappa_{N} \over \kappa_{N+1}}
                 \right) .
\label{III_J3}
\end{equation}
Equating (\ref{III_J2}) and (\ref{III_J3}) and eliminating $ l_{N} $ 
using (\ref{ops_l}) gives (\ref{III_rRecur:a}). To deduce (\ref{III_rRecur:b}) 
we apply an analogous strategy to
\begin{equation}
  J_{2} := \int_{\EuScript{C}} {dz \over 2\pi iz} 
       w'(z)\phi_{N+1}(z)\overline{\phi_{N}(z)}.
\end{equation}
\end{proof}

To deduce (\ref{AvM_III:a}), (\ref{AvM_III:b}) from (\ref{III_rRecur:a}),
(\ref{III_rRecur:b}) we first subtract (\ref{III_rRecur:a}) from (\ref{III_rRecur:b})
to find
\begin{equation}
   \half\sqrt{t}\left( L_{N+1}+L_{N} \right)+2\mu = 0, \qquad
   L_{N} := \bar{r}_{N}r_{N-1}-r_{N}\bar{r}_{N-1} . 
\label{III_Rdiff}
\end{equation}
Using (\ref{III_rInitial}) and a Bessel function identity this equation is to be 
solved subject to the initial condition
\begin{equation}
   L_{1} := -{2\mu \over \sqrt{t}} .
\end{equation}
It follows that the solution of (\ref{III_Rdiff}) is the constant
\begin{equation}
   L_{N} = -{2\mu \over \sqrt{t}}, \qquad N=1,2,3,\ldots
\end{equation}
and thus
\begin{equation}
   \half\sqrt{t}\left( \bar{r}_{N+1}r_{N}-r_{N+1}\bar{r}_{N} \right)
   + \mu = 0 .
\label{III_rRecur:c}
\end{equation}
Using this to substitute for $ \mu $ in (\ref{III_rRecur:a}) and (\ref{III_rRecur:b})
gives (\ref{AvM_III:a}) and (\ref{III_rRecur:b}) respectively. Furthermore 
(\ref{III_rRecur:c}) can be summed to give
\begin{equation}
   \half\sqrt{t}\left( {\bar{l}_{N}\over \kappa_{N}}-{l_{N}\over \kappa_{N}} \right)
   + \mu N = 0 .
\label{III_lRecur}
\end{equation}

We turn our attention now to formulae for $ \tau^{\rm III'}[N] $, $r_N $ and 
$ \bar{r}_N $ in terms of generalised hypergeometric functions. From earlier work
\cite{Ka_1993}, \cite{FW_2002a} we know
\begin{equation}
     \tau^{\rm III'}[N](t;\mu) 
    = \left({ \sqrt{t} \over 2}\right)^{N\mu}
      \prod^{N}_{j=1}{j! \over \Gamma(j+\mu)}
      {}^{\vphantom{(1)}}_{0}F^{(1)}_{1}(;N+\mu;t_1,\ldots,t_N)|_{t_1=\ldots =t_N=t/4}.
\label{III_0F1}
\end{equation}
It follows from this and (\ref{III_IToep}), (\ref{III_tau}) and (\ref{ops_Refl})
that
\begin{align}
   r_{N} & = (-1)^N \left({\sqrt{t} \over 2}\right)^{N}{1 \over (\mu+1)_N}
   { {}^{\vphantom{(1)}}_{0}F^{(1)}_{1}(;N+1+\mu;t_1,\ldots,t_N) \over 
     {}^{\vphantom{(1)}}_{0}F^{(1)}_{1}(;N+\mu;t_1,\ldots,t_N) }\Big|_{t_1=\ldots =t_N=t/4},
   \\
   \bar{r}_{N} & = (-1)^N \left({2 \over \sqrt{t}}\right)^{N}(\mu)_N
   { {}^{\vphantom{(1)}}_{0}F^{(1)}_{1}(;N-1+\mu;t_1,\ldots,t_N) \over 
     {}^{\vphantom{(1)}}_{0}F^{(1)}_{1}(;N+\mu;t_1,\ldots,t_N) }\Big|_{t_1=\ldots =t_N=t/4}.
\end{align}
Note that the small-$t$ expansions are more forthcoming from these formulae
than from the Toeplitz determinants.

Our final point in relation to $ \tau^{\rm III'}[N] $ concerns the known 
recurrence scheme for $ \tau^{\rm III'}[N] $ in terms of the variables 
$ p_N, q_N $ specifying the corresponding Hamiltonian in the Painlev\'e systems
approach to \PIIIa. The idea here is that we start from the Hamiltonian for the 
\PIIIa system
\begin{equation}\label{2.20a}
  tH_n^{\rm III'} 
  = q_n^2 p_n^2 - (q_n^2+v_1q_n -t) p_n + \half(v_1+v_2)q_n ,
\end{equation}
with special parameters $(v_1,v_2) = (v^{(0)}_1,v^{(0)}_2) = (\mu, -\mu) $ and 
corresponding special values $ p = p_0 = 0, q = q_0 $ for some particular $ q_0 $ 
(see e.g. \cite{FW_2002a}). A sequence of Hamiltonians is constructed from this 
seed by application of the Schlesinger transformation $ T_1 $ with the action 
on the parameters $ T_1\cdot (v_1,v_2) = (v_1+1,v_2+1) $ and some 
explicit actions on $ p $ and $ q $ involving rational functions of 
$ p|_{(v_1,v_2)} $ and $ q|_{(v_1,v_2)} $. Thus we set 
\begin{equation}
  tH_n^{\rm III'} := 
  tH^{\rm III'}\Big|_{T_1^n (v^{(0)}_1,v^{(0)}_2)} ,
\end{equation}
and introduce the corresponding $ \tau$-function $ \tau^{\rm III'}_n $ by the 
requirement that 
\begin{equation}
  H_n^{\rm III'} := {d \over dt}\log \tau^{\rm III'}_n .
\label{III_tauDef}
\end{equation}
We know from \cite{FW_2002b} that with $ (v^{(0)}_1,v^{(0)}_2) = (\mu, -\mu) $
the sequence $ \{ \tau^{\rm III'}_n \}_{n=0,1,2,\ldots} $ is realised by
\begin{equation}
  \tau^{\rm III'}_n = t^{-n\mu/2}\tau^{\rm III'}[n](t;\mu)\Big|_{t \mapsto 4t} .
\end{equation}
With $ p_n := T_1^n p|_{(v^{(0)}_1,v^{(0)}_2)} $
and $ q_n := T_1^n q|_{(v^{(0)}_1,v^{(0)}_2)} $ further 
theory associated with $ T_1 $ led to the following recurrence scheme.

\begin{proposition}[\cite{FW_2003a}]\label{III_hProp}
We have 
\begin{align}
  {\tau^{\rm III'}[N+1] \tau^{\rm III'}[N-1] \over
  (\tau^{\rm III'}[N])^2} \Big |_{t \mapsto 4t} & = p_N, \quad (N = 1,2,\ldots)
  \label{III_hRecur:c} \\
  p_{N+1} & = {q_N^2 \over t} (p_N - 1) - \mu{q_N \over t} + 1
  \quad (N = 0,1,\ldots)
  \label{III_hRecur:a} \\
  q_{N+1} & = - {t \over q_N}
              + {(N+1) t \over q_N[q_N(p_N-1) - \mu] + t}
  \quad (N = 0,1,\ldots)
  \label{III_hRecur:b}
\end{align}
subject to the initial conditions
\begin{gather}
  p_0 = 0, \qquad 
  q_0 = t{d \over dt}\log t^{-\mu/2}I_\mu(\sqrt{t}) \Big|_{t \mapsto 4t} \\
  \tau^{\rm III'}[0] = 1, \qquad 
  \tau^{\rm III'}[1]\big|_{t \mapsto 4t} = I_\mu(2\sqrt{t}).
\end{gather}
\end{proposition}

Our interest is in the relationship between the variables $ p_N, q_N $ in these
recurrences and the reflection coefficients of Prop.\ref{III_rProp}. This is
given by the following result.
\begin{proposition}\label{III_Xfm}
The Hamiltonian variables $ q_N, p_N $ in Prop.\ref{III_hProp} are related to 
the reflection coefficients in Prop.\ref{III_rProp} by 
\begin{align}
   p_N & = 1-r_N\bar{r}_N|_{t \mapsto 4t},
   \label{IIIH_Ops:b} \\
   q_N & = -\sqrt{t}{r_{N+1} \over r_N} \Big|_{t \mapsto 4t}.
   \label{IIIH_Ops:a}
\end{align}
\end{proposition}
\begin{proof}
The equation (\ref{IIIH_Ops:b}) follows immediately upon comparing (\ref{III_hRecur:c})
with (\ref{ops_I0}). Substituting (\ref{IIIH_Ops:b}) in (\ref{III_hRecur:a})
we see that (\ref{III_rRecur:c}) results if we also substitute for $ q_N $ according
to (\ref{IIIH_Ops:a}). Furthermore, if we combine (\ref{III_hRecur:a}) and 
(\ref{III_hRecur:b}) into
\begin{equation}
   q_N + {t\over q_{N-1}} = {N \over p_N},
\end{equation}
we see that the substitutions (\ref{IIIH_Ops:a}) and (\ref{IIIH_Ops:b}) in this 
equation yields (\ref{AvM_III:a}). 
\end{proof}

Thus the structure of the recurrences is such that the transformation equations
(\ref{IIIH_Ops:a}) and (\ref{IIIH_Ops:b}) can essentially be determined by 
inspection. However when we come to study the analogous recurrences for the 
\PV $ \tau$-function (\ref{V_Toep2}) this is not possible and a more systematic
procedure is called for. With this in mind, let us then present a more
systematic approach to the derivation of (\ref{IIIH_Ops:a}).

For this purpose, in addition to the shift operator $ T_1 $ which increments $ N $ 
in $ \tau^{\rm III'}[N] $, we introduce the other fundamental Schlesinger 
transformation of the \PIIIa system $ T_2 $ with the action on the parameters
$ T_2\cdot(v_1,v_2) = (v_1+1,v_2-1) $. Recalling that in $ \tau^{\rm III'}[N] $
$ (v^{(0)}_1,v^{(0)}_2) = (\mu,-\mu) $, this operator then increments $ \mu $ by
unity. From \cite{FW_2002a} we know that for general parameters
\begin{equation}
   T_1\cdot tH^{\rm III'} = tH^{\rm III'}-q(p-1), \quad 
   T_2\cdot tH^{\rm III'} = tH^{\rm III'}-qp,
\end{equation}
and thus in particular
\begin{align}
   q_N & = t{d \over dt}\log{T_1\cdot \tau_N \over T_2\cdot \tau_N}
   \\
       & = \half(N-\mu) - t{d \over dt}\log\kappa^2_N r_N\Big|_{t \mapsto 4t},
\label{III_qxfm}
\end{align}
where to obtain the second equality use has been made of (\ref{III_tauDef}),
(\ref{ops_Refl}) and (\ref{ops_I0kappa}).
To determine the $t$-derivatives of the orthogonal polynomial coefficients
we find the $t$-derivatives of the polynomials themselves. Differentiating
the orthonormality condition (\ref{ops_onorm}) with $ w(z) $ given by 
(\ref{III_wgt}) we have
\begin{equation}
  0 = -{\dot{w_0} \over w_0}\delta_{mn}
    + \int_{\mathbb{T}} {dz \over 2\pi iz} 
      \tilde{w}[\dot{\phi}_m+\half z\phi_m]\bar{\phi}_n
    + \int_{\mathbb{T}} {dz \over 2\pi iz} 
      \tilde{w}\phi_m[\bar{\dot{\phi}}_n+\half \overline{z\phi}_n] ,
\label{III_normDer}
\end{equation}
where $ \dot{} $ represents differentiation with respect to $ t $.
Now $ \dot{\phi}_n+\half z\phi_n $ is of degree $ n+1 $ in $ z $ and has no 
components in $ \phi_m $ for $ m \geq n+2 $. With this established it follows
from (\ref{III_normDer}) that $ \dot{\phi}_n+\half z\phi_n $ has no components
in $ \phi_m $ for $ m \leq n-2 $ either and so
\begin{equation}
    \dot{\phi}_n+\half z\phi_n 
  = \bar{a}_n\phi_{n+1}+\bar{b}_n\phi_{n+1}+\bar{c}_n\phi_{n-1}.
\label{III_opsDer}
\end{equation} 
Equating coefficients of the highest monomial in (\ref{III_opsDer}) and recalling
(\ref{ops_coeff}) gives $ \bar{a}_n = \kappa_n/2\kappa_{n+1} $ while (\ref{III_normDer}) 
in the case $ m=n-1 $ yields $ \bar{c}_n = -\kappa_{n-1}/2\kappa_n $. 
Finally, setting $ m=n $ in (\ref{III_normDer}) and recalling (\ref{Ibessel})
shows
\begin{equation}
   \bar{b}_n+b_n = {\dot{I}_{\mu} \over I_{\mu}}.
\end{equation}
Substituting for $ \bar{a}_n $ and $ \bar{c}_n $ in (\ref{III_opsDer}), and 
using the three-term recurrence (\ref{ops_ttr:a}) to substitute for 
$ \bar{a}_n\phi_{n+1}-\half z\phi_n $ we deduce
\begin{equation}
   \dot{\phi}_n = \left[ \bar{b}_n + \half{r_{n+1} \over r_n} \right]\phi_n
       -\half{\kappa_{n-1}\over \kappa_n}\left[ 1+{r_{n+1} \over r_n}z \right]\phi_{n-1}.
\end{equation}
Recalling (\ref{ops_coeff}) it follows that
\begin{align}
   {\dot{\kappa}^2_n \over \kappa^2_n} 
  & = {\dot{I}_{\mu} \over I_{\mu}} + \half(r_{n+1}\bar{r}_n+\bar{r}_{n+1}r_{n})
  \\
   {\dot{r}_n \over r_n}
  & = \half(r_{n+1}-r_{n-1})\left({1\over r_{n}}-\bar{r}_n\right)
  \\ 
   {\dot{\bar{r}}_n \over \bar{r}_n}
  & = \half(\bar{r}_{n+1}-\bar{r}_{n-1})\left({1\over \bar{r}_n}-r_{n}\right).
\end{align} 
Making use of the first two of these relations in (\ref{III_qxfm}) and simplifying 
reclaims (\ref{IIIH_Ops:a}).
 
\section{The \PV System}
\setcounter{equation}{0}

The weight function
\begin{equation}
   w(z) = (1+z)^{\mu}(1+1/z)^{\nu} e^{tz} ,
\label{V_wgt}
\end{equation}
is analytic in the cut plane $ z \in \CC\backslash(-\infty,-1] $. For 
$ \Re(\mu+\nu+1) > 0 $ the singularity is integrable at $ z=-1 $ so with this
restriction there is no need to deform $ \mathbb{T} $ in (\ref{ops_Fourier}).
The transition from the \PV to the \PIIIa system can be achieved by making the
replacements 
\begin{equation}
  \nu \mapsto \nu-\mu,\qquad z_l \mapsto 2\nu z_l/\sqrt{t},\qquad t \mapsto t/4\nu
\label{V_to_III}
\end{equation}
with $ t, N, \mu $ fixed and then taking the limit $ \nu \to \infty $. In this way
the Toeplitz determinant (\ref{V_Toep}) reduces to (\ref{III_Toep}). 
In \cite{FW_2003a} we have shown that with $ w $ given by (\ref{V_wgt})
\begin{equation}
   \int_{\mathbb{T}} {dz \over 2\pi iz}z^{-n}w(z)
   = {\Gamma(\mu+\nu+1) \over \Gamma(\mu+1-n)\Gamma(\nu+1+n)}
           {}_1F_1(-n-\nu;\mu-n+1;-t), \quad n \in \ZZ ,
\label{V_toepM}
\end{equation}
and thus $ I^{\epsilon}_N[w] $ as specified by (\ref{ops_Idefn}) can be made
explicit. In particular it follows from (\ref{V_toepM}), (\ref{ops_Idefn})
and (\ref{ops_Refl}) that 
\begin{equation}
   r_{1} = -{\nu \over \mu+1}{ {}_1F_1(-\nu+1;\mu+2;-t) \over {}_1F_1(-\nu;\mu+1;-t) },
   \quad
   \bar{r}_{1} = -{\mu \over \nu+1}{ {}_1F_1(-\nu-1;\mu;-t) \over {}_1F_1(-\nu;\mu+1;-t) }.
\label{V_rInitial}
\end{equation}

As with the weight (\ref{III_wgt}), recurrences for the reflection coefficients
$ r_N, \bar{r}_N $ in the case of the weight (\ref{V_wgt}) can be deduced from the
work of Adler and van Moerbeke \cite{AvM_2002}. Thus in their first case one specialises
their parameters
$ \gamma''_1 = \gamma'_2 = \gamma''_2 = d_2 = 0, d_1 = -1, \gamma = -\nu, 
  \gamma'_1 = \mu+\nu $ which gives their $ (a,b,c) = (1,1,0) $, and one sets
$ P_1(z) = tz, P_2(z) = 0 $.
With the identification $ x_N = r_{N}, y_N = \bar{r}_{N} $, from \cite{AvM_2002}
eq.(0.0.14) we then read off the recurrence
\begin{equation}
   - tr_{N+1}\bar{r}_{N} + tr_{N}\bar{r}_{N-1} 
   + (N+1+\nu)r_{N}\bar{r}_{N+1} - (N-1+\nu)r_{N-1}\bar{r}_{N} = 0 ,
\label{V_AvM_1}
\end{equation}
which is of order $ 2/2 $. Also \cite{AvM_2002} eq.(0.0.15) reads 
\begin{multline}
   v_{N+1}\left[ N+1+\nu + tr_{N+2}\bar{r}_{N} \right] 
   - v_{N}\left[ N+\nu + tr_{N+1}\bar{r}_{N-1} \right] + r_{N+1}\bar{r}_{N} \\
  = v_{1}(1+\nu+tr_{2}\bar{r}_{0}) + r_{1}\bar{r}_{0} ,
\label{V_AvM_2a}
\end{multline}
where $ v_N := 1-r_{N}\bar{r}_{N} $ which is also of order $ 2/2 $. The right
hand side of (\ref{V_AvM_2a}) can be simplified. First we note that it can be
written in terms of the Fourier components of (\ref{V_wgt}) according to
\begin{equation}
  (\nu+1)(1-w_{-1}w_{1})+t(w^2_{-1}-w_{-2})-w_{-1}
\end{equation}
and thus in the light of (\ref{V_toepM}) in terms of the confluent hypergeometric 
function. Utilising the contiguous relation for the confluent hypergeometric function 
$ {}_1F_1(a;b;-t) $, \cite{AS_1970} (13.4.7, p.507) in the two cases 
$ a=-\nu, b=\mu+1 $ and $ a=-\nu+1, b=\mu+2 $ one can show this is precisely unity,
and so (\ref{V_AvM_2a}) simplifies to
\begin{equation}
   v_{N+1}\left[ N+1+\nu + tr_{N+2}\bar{r}_{N} \right] 
   - v_{N}\left[ N+\nu + tr_{N+1}\bar{r}_{N-1} \right] + r_{N+1}\bar{r}_{N} = 1. 
\label{V_AvM_2}
\end{equation}

We can use an orthogonal polynomial approach to derive recurrences for 
$ r_{N},\bar{r}_{N} $, and from equations used in the derivation (\ref{V_AvM_1})
and (\ref{V_AvM_2}) can be reclaimed.
\begin{proposition}\label{V_rProp}
The reflection coefficients $ r_{N}, \bar{r}_{N} $ for the orthogonal polynomial
system on the unit circle with the weight (\ref{V_wgt}) satisfy the coupled system 
of $ 2/1 $ and $ 1/2 $ order recurrence relations 
\begin{align}
   (N-1+\nu)r_{N-1} + (N+\mu+t)r_{N} & = 
   -t(1-r_{N}\bar{r}_{N})r_{N+1} + t r^2_{N}\bar{r}_{N-1},
   \label{V_rRecur:a}\\
   (N+1+\nu)\bar{r}_{N+1} + (N+\mu+t)\bar{r}_{N} & = 
   -t(1-r_{N}\bar{r}_{N})\bar{r}_{N-1} + t \bar{r}^2_{N}r_{N+1},
   \label{V_rRecur:b}
\end{align}
subject to the initial conditions (\ref{V_rInitial}).
\end{proposition}

\begin{proof}
As with the derivation of the recurrences of Proposition \ref{III_rProp} we will
take the Freud approach, but now working with two integrals for each recurrence
rather than the single integral for each recurrence required to derive the 
recurrences of Proposition \ref{III_rProp}. The first integral we consider is 
\begin{equation}
  J_1 := 
    \int_{\TT} {dz \over 2\pi iz} (1+z)w'(z)\phi_{N}(z)\overline{\phi_{N}(z)} .
\label{V_J1}
\end{equation}
Integrating this by parts, employing (\ref{ops_prod}) and the orthogonality 
condition shows
\begin{equation}
   J_1 := N{\bar{l}_{N} \over \kappa_{N}} - (N+1){\bar{l}_{N+1} \over \kappa_{N+1}}.
\label{V_int1_rhs}
\end{equation}
On the other hand we note from (\ref{V_wgt}) that
\begin{equation}
   {w' \over w} = {\mu+\nu \over 1+z} - {\nu \over z} + t,
\end{equation}
substituting this in (\ref{V_J1}), an analogous calculation shows
\begin{equation}
   J_1 = \mu 
       - \nu\left( {\bar{l}_{N} \over \kappa_{N}}-{\bar{l}_{N+1} \over \kappa_{N+1}}
            \right)
       + t\left( {l_{N} \over \kappa_{N}}-{l_{N+1} \over \kappa_{N+1}} \right) + t.
\label{V_int1_lhs}
\end{equation}
Upon employing (\ref{ops_l}) one can equate (\ref{V_int1_rhs}) and (\ref{V_int1_lhs})
and solve for $ \bar{l}_{N+1} $ to obtain
\begin{equation}
   {\bar{l}_{N+1} \over \kappa_{N+1}} = 
   -\mu - t(1-\bar{r}_{N}r_{N+1}) - (N+\nu)r_{N}\bar{r}_{N+1}.
\label{V_int1}
\end{equation}
Note that one could perform another differencing at this point and eliminate 
$ \bar{l}_{N} $ in favour of $ r_{N}, \bar{r}_{N} $ but then $ \mu $ would disappear 
from the ensuing relations. This is a clear indication that the recurrence system 
would be raised unnecessarily in order, so we seek another relation for 
$ \bar{l}_{N} $, and this is found by considering the following integral
\begin{equation}
  J_2 := 
    \int_{\TT} {dz \over 2\pi iz} z(1+z)w'(z)\phi_{N}(z)\overline{\phi_{N+1}(z)}.
\end{equation}
The methods of evaluation used to derive (\ref{V_int1_rhs}) and (\ref{V_int1_lhs})
now yield
\begin{equation}
  J_2 = (N+1){\kappa_{N+1} \over \kappa_{N}} - (N+1){\kappa_{N} \over \kappa_{N+1}} 
   - {\bar{l}_{N+1} \over \kappa_{N}},
\label{V_int2_rhs}
\end{equation}
and
\begin{equation}
   J_2 = \mu{\kappa_{N} \over \kappa_{N+1}}
 + t\left( {\kappa_{N} \over \kappa_{N+1}} + {l_{N} \over \kappa_{N+1}}
           - {l_{N+2} \over \kappa_{N+2}}{\kappa_{N} \over \kappa_{N+1}} \right),
\label{V_int2_lhs}
\end{equation}
respectively. Equating (\ref{V_int2_rhs}) and (\ref{V_int2_lhs}), and again 
employing (\ref{ops_l}), it follows
\begin{equation}
   {\bar{l}_{N+1} \over \kappa_{N+1}} = 
   N+1 - \left[ N+1+\mu+t - t(r_{N+2}\bar{r}_{N+1}+r_{N+1}\bar{r}_{N}) \right]
              (1-r_{N+1}\bar{r}_{N+1}).
\label{V_int2}
\end{equation}
Equating (\ref{V_int1}) and (\ref{V_int2}) gives (\ref{V_rRecur:a}).
The second recurrence can be found in a similar manner by eliminating $ l_{N+1} $ from 
expressions arising from evaluation of the integrals
\begin{equation*}
  J_3 := 
    \int_{\TT} {dz \over 2\pi iz} (1+z)w'(z)\phi_{N+1}(z)\overline{\phi_{N}(z)},
\end{equation*}
and
\begin{equation*}
  J_4 :=
    \int_{\TT} {dz \over 2\pi iz} z(1+z)w'(z)\phi_{N}(z)\overline{\phi_{N}(z)}.
\end{equation*}
The two expressions for $ l_{N+1} $ are respectively
\begin{align}
   {l_{N+1} \over \kappa_{N+1}}
  = & (N+1)r_{N+1}\bar{r}_{N+1} - (\nu+tr_{N+2}\bar{r}_{N})(1-r_{N+1}\bar{r}_{N+1})
  \label{V_int3} \\
  = & -\nu - (N+\mu+t)r_{N+1}\bar{r}_{N} - t(1-r_{N}\bar{r}_{N})r_{N+1}\bar{r}_{N-1}
  \nonumber \\
    & - t(1-r_{N+1}\bar{r}_{N+1})r_{N+2}\bar{r}_{N} + tr^2_{N+1}\bar{r}^2_{N},
  \label{V_int4}
\end{align}
and this yields (\ref{V_rRecur:b}).
\end{proof}

Let us now show how we can derive the recurrences of Adler and van Moerbeke
(\ref{V_AvM_1}), (\ref{V_AvM_2}) from the workings of the proof of Proposition 
\ref{V_rProp}. As already remarked, subtracting from (\ref{V_int1}) the same 
equation with $ N $ replaced by $ N-1 $ and recalling (\ref{ops_l}) gives a 
difference equation in $ r_N, \bar{r}_N $. This difference equation is in fact 
precisely (\ref{V_AvM_1}). Note that with $ \Delta_N $ the forward difference
operator with respect to $ N $, (\ref{V_int1}) itself can be written
\begin{equation}
   \Delta_N [(N+\nu){\bar{l}_N\over \kappa_N}-t{l_N\over \kappa_N}] = -\mu-t .
\end{equation}
Summing this over $ N $ shows
\begin{equation}
   (N+\nu){\bar{l}_N\over \kappa_N}-t{l_N\over \kappa_N}= -(\mu +t)N ,
\label{V_lReln}
\end{equation}
which then is the basic relation underlying (\ref{V_AvM_1}). The recurrence 
(\ref{V_AvM_2}) is obtained by subtracting from (\ref{V_int3}) the same equation
with $ N $ replaced by $ N-1 $, and recalling (\ref{ops_l}).

We consider next formulae for $ \tau^{\rm V}[N](t;\mu,\nu) $, $ r_N, \bar{r}_N $
in terms of generalised hypergeometric functions. From earlier work \cite{Ka_1993},
\cite{FW_2002b} we know
\begin{equation}
  \tau^{\rm V}[N](t;\mu,\nu)
  = \prod^{N-1}_{j=0}{\Gamma(\mu+\nu+1+j)\Gamma(1+j) \over 
                      \Gamma(\mu+1+j)\Gamma(\nu+1+j)}
     {}^{\vphantom{(1)}}_{1}F^{(1)}_{1}(-\nu;N+\mu;t_1,\ldots,t_N)
                    \Big|_{t_1=\ldots =t_N=-t} .
\label{V_1F1}
\end{equation}
Noting from (\ref{ops_Idefn}), (\ref{ops_Uint}) and (\ref{V_modulus}) that 
for the weight (\ref{V_wgt})
\begin{equation}
  I^1_N[w] = \tau^{\rm V}[N](t;\mu+1,\nu-1), \quad
  I^{-1}_N[w] = \tau^{\rm V}[N](t;\mu-1,\nu+1) ,
\end{equation}
it follows from this and (\ref{ops_Refl}) that
\begin{align}
   r_{N} & = (-1)^N {(\nu)_{N} \over (\mu+1)_{N}} 
   { {}^{\vphantom{(1)}}_{1}F^{(1)}_{1}(-\nu+1;N+1+\mu;t_1,\ldots,t_N) \over 
     {}^{\vphantom{(1)}}_{1}F^{(1)}_{1}(-\nu;N+\mu;t_1,\ldots,t_N) }
                    \Big|_{t_1=\ldots =t_N=-t},
   \label{V_genH:a} \\
   \bar{r}_{N} & = (-1)^N {(\mu)_{N} \over (\nu+1)_{N}} 
   { {}^{\vphantom{(1)}}_{1}F^{(1)}_{1}(-\nu-1;N-1+\mu;t_1,\ldots,t_N) \over 
     {}^{\vphantom{(1)}}_{1}F^{(1)}_{1}(-\nu;N+\mu;t_1,\ldots,t_N) }
                    \Big|_{t_1=\ldots =t_N=-t}.
   \label{V_genH:b}
\end{align}
An immediate consequence of (\ref{V_genH:a}) and (\ref{V_genH:b}) is the explicit
values at $ t = 0 $,
\begin{gather}
  r_{N} = (-1)^N{(\nu)_{N}\over (\mu+1)_{N}} ,\quad
 \bar{r}_{N} = (-1)^N{(\mu)_{N}\over (\nu+1)_{N}} ,\\
  {l_{N}\over \kappa_N} = -{N\nu\over N+\mu} ,\quad 
  {\bar{l}_{N}\over \kappa_N} = -{N\mu\over N+\nu} .
\end{gather}
Again we see that the transition from the generalised hypergeometric function in 
the \PV system (\ref{V_1F1}) to that of the \PIIIa system (\ref{III_0F1}) is 
facilitated by making the replacements (\ref{V_to_III}) and taking the limit 
$ \nu \to \infty $
\begin{multline}
  {}^{\vphantom{(1)}}_{1}F^{(1)}_{1}(\sigma-\nu;N+\mu;t_1,\ldots,t_N)
  \Big|_{t_1=\ldots =t_N=-t/4\nu} \\
  \mathop{\rightarrow}\limits_{\nu \to \infty}
  {}^{\vphantom{(1)}}_{0}F^{(1)}_{1}(;N+\mu;t_1,\ldots,t_N)
  \Big|_{t_1=\ldots =t_N=t/4} ,
 \label{V-III_1F1}
\end{multline}
for all fixed $ \sigma, N+\mu, t $. Thus the corresponding limiting forms for the 
reflection coefficients are
\begin{equation}
 r^{\rm V}_{N} 
 \mathop{\sim}\limits_{\nu \to \infty}
  \left({2\nu \over \sqrt{t}}\right)^{N} r^{\rm III'}_N ,
 \qquad
 \bar{r}^{\rm V}_{N} 
 \mathop{\sim}\limits_{\nu \to \infty}
  \left({\sqrt{t} \over 2\nu}\right)^{N} \bar{r}^{\rm III'}_N ,
 \label{V-III_r}   
\end{equation}
where we distinguish the two systems only when some confusion could arise.
Another consequence of this limiting process is that the recurrences in 
$ r_N, \bar{r}_N $ (\ref{V_rRecur:a}), (\ref{V_rRecur:b}) reduce to a
sum of (\ref{III_rRecur:a}) and (\ref{III_rRecur:c}), and of 
(\ref{III_rRecur:b}) and (\ref{III_rRecur:c}) respectively. In addition we find
that (\ref{V_lReln}) reduces to (\ref{III_lRecur}). 

It remains to compare the recurrences (\ref{V_rRecur:a}), (\ref{V_rRecur:b})
determining $ \tau^{\rm V}[N] $ through the recurrence (\ref{ops_I0}), and 
recurrences satisfied by the Hamiltonian variables $ q_N, p_N $ in the Painlev\'e
systems approach to \PV which also determine $ \tau^{\rm V}[N] $ through a 
recurrence \cite{FW_2003a}. In relation to the latter, let $ v_1+v_2+v_3+v_4 = 0 $
and introduce the Hamiltonian
\begin{multline}
  tH^{\rm V}
  = q(q-1)p(p+t) - (v_2-v_1+v_3-v_4)q p + (v_2-v_1)p + (v_1-v_3)tq .
\label{2.15V}
\end{multline}
Starting with a special solution when $ v_3-v_4 = v^{(0)}_3-v^{(0)}_4 = 0 $,
$ q = q_0 = 1 $, $ p = p_0 $ for some particular $ p_0 $ (see \cite{FW_2003a}),
a sequence of Hamiltonians is constructed by application of the Schlesinger 
transformation $ T^{-1}_0 $ with the action on the parameters 
$ T^{-1}_0 \cdot (v_1,v_2,v_3,v_4) = 
  (v_1-\quarter,v_2-\quarter,v_3-\quarter,v_4+\tquarter) $ to obtain 
\begin{equation}
  tH_n^{\rm V} = tH^{\rm V}\Big|_{v \mapsto
      (v_1-n/4,v_2-n/4,v_3-n/4,v_4+3n/4)} .           
\end{equation}
The corresponding sequence of $ \tau $-functions $ \tau^{\rm V}_n $ are specified
so that 
\begin{equation}
  H_n^{\rm V} = {d \over dt}\log \tau^{\rm V}_n .
\label{V_tauDefn}
\end{equation}
We know from \cite{FW_2003a} that with 
$ v^{(0)}_1-v^{(0)}_2=\mu, v^{(0)}_1-v^{(0)}_3=-\nu, v^{(0)}_3-v^{(0)}_4=0 $ the
sequence $ \{\tau^{\rm V}_n\}_{n=0,1,\ldots} $ is realised by
\begin{equation}\label{2.30V}
  e^{\nu t}\tau_n^{\rm V} 
  = \left[{\Gamma(\mu+1) \over \Gamma(\mu+\nu+1)}\right]^n
    \prod^{n-1}_{l=0}\Gamma(\nu+l+1)\tau^{\rm V}[n](t;\mu,\nu) .
\end{equation}
Furthermore $ \{ \tau^{\rm V}[N] \}_{N=2,3,\ldots} $ is determined by the following
recurrence scheme \cite{FW_2003a}.

\begin{proposition}
Let
\begin{equation}
   x_N = (p_N+t)q_N + \half\mu, \qquad y_N = {1 \over q_N} .
\label{V_xyDefn}
\end{equation}
The sequences $ \{ \tau^{\rm V}[N] \}_{N=0,1,\ldots} $, $\{x_N\}_{N=0,1,\dots}$, 
$\{y_N\}_{N=0,1,\dots}$ satisfy the coupled recurrences
\begin{align}
  (N+\nu){\tau^{\rm V}[N+1]\tau^{\rm V}[N-1] \over (\tau^{\rm V}[N])^2}
  & = \left( x_N-{t \over y_N}-\nu-\half\mu \right)
      \left( {1 \over y_N}-1 \right)+N
  \label{V_hRecur:c} \\
  x_N + x_{N-1}
  & = {t \over y_N} - {N \over 1 - y_N}
  \label{V_hRecur:a} \\
  y_N y_{N+1} 
  & = t {x_N +N+1+\nu+\half\mu \over x_N^2 - \quarter\mu^2},
  \label{V_hRecur:b}
\end{align}
subject to the initial conditions
\begin{gather}
  x_0 = t+\half\mu+t{d \over dt}\log {}_1F_1(-\nu;\mu+1;-t), \quad
  y_0 = 1 
  \label{V_hInit:a} \\
  \tau^{\rm V}[0] = 1, \quad 
  \tau^{\rm V}[1] = {\Gamma(\mu+\nu+1) \over \Gamma(\mu+1)\Gamma(\nu+1)}
                    {}_1F_1(-\nu;\mu+1;-t) .
  \label{V_hInit:c}
\end{gather}
\end{proposition}

By adopting a strategy analogous to the method of derivation of (\ref{IIIH_Ops:a}), 
(\ref{IIIH_Ops:b}) given below the proof of Proposition \ref{III_Xfm},
the relationship between the variables $ x_N, y_N $ and $ r_N, \bar{r}_N $ can
be deduced.
\begin{proposition}\label{V_Xfm}
The Hamiltonian variables $ x_N, y_N $ and the reflection coefficients 
$ r_N, \bar{r}_N $ are related by the equations
\begin{gather}
   \left({1 \over y_N}-1\right)\left[ x_N-{t \over y_N}-\nu-\half\mu \right]+N
  = (N+\nu)(1-r_N\bar{r}_N)
 \label{VH_Ops:a} \\
   \left[ x_N-{t \over y_N}-\nu-\half\mu \right]
   \left( 1+{N+\nu \over 
             \left(x_N-{\displaystyle t \over \displaystyle y_N}-\half\mu\right)(1-y_N)-\nu}
   \right)-N
 \nonumber\\
  = t{r_{N+1} \over r_{N}}(1-r_N\bar{r}_N)
 \label{VH_Ops:b} \\
   (1-y_N)\left[ x_N-{t \over y_N}+\half\mu \right]+N
  = -t{\bar{r}_{N-1} \over \bar{r}_{N}}(1-r_N\bar{r}_N) .
 \label{VH_Ops:c}
\end{gather}
Consequently
\begin{gather}
  y_N = {\nu+tr_{N+1}\bar{r}_N \over \bar{r}_N[tr_{N+1}+(N+\nu)r_{N}]}
      \label{V_xfm:a} \\
  x_N-{t \over y_N}-\half\mu = -tr_{N+1}\bar{r}_N .
      \label{V_xfm:b}
\end{gather}
\end{proposition}

\begin{proof}
The equation (\ref{VH_Ops:a}) follows immediately upon substituting for 
$ \tau^{\rm V}[N+1]\tau^{\rm V}[N-1]/(\tau^{\rm V}[N])^2 $ according to the 
right-hand side of the general relation (\ref{ops_I0}) in (\ref{V_hRecur:c}).
For the remaining two equations, adapting the method of derivation of 
(\ref{IIIH_Ops:a}), (\ref{IIIH_Ops:b}) given below the proof of Proposition 
\ref{III_Xfm}, we require a shift operator that has the action
$ \mu \mapsto \mu+1, \nu \mapsto \nu-1 $ and leaves $ N $ fixed. Such an operator 
is the Schlesinger transformation $ T^{-1}_2 $ with the action on the parameters
$ T^{-1}_2\cdot(v_1,v_2,v_3,v_4) 
             = (v_1+\tquarter,v_2-\quarter,v_3-\quarter,v_4-\quarter) $
which was studied in \cite{FW_2002a}. From the explicit form of $ T^{-1}_2 $ in 
terms of operators associated with the root lattice $ A_3 $, and the actions of
these operators on the Hamiltonian and associated variables in \cite{FW_2002a}, 
we can compute that
\begin{align}
  T^{-1}_2\cdot tH^{\rm V}
  & = tH^{\rm V} + \left(q+{v_1-v_3 \over p}\right)
         \left( p-{v_1-v_4 \over q-1+{\displaystyle v_1-v_3 \over \displaystyle p}}
         \right)+t+v_3-v_4
  \\
  T_2\cdot tH^{\rm V} 
  & = tH^{\rm V} + (q-1)\left(p-{v_2-v_1 \over q}\right)-t-(v_3-v_4)
\end{align}
Recalling (\ref{V_tauDefn}), (\ref{2.30V}) and (\ref{V_xyDefn}) it follows that
\begin{gather}
   \left[ x_N-{t \over y_N}-\nu-\half\mu \right]
   \left( 1+{N+\nu \over 
             \left(x_N-{\displaystyle t \over \displaystyle y_N}-\half\mu\right)(1-y_N)-\nu}
   \right)-N
  = t{d \over dt}\log r_n
  \label{V_rxy:a} \\
  \begin{split}
   (1-y_N)\left[ x_N-{t \over y_N}+\half\mu \right]+N
  = t{d \over dt}\log \bar{r}_n.
  \end{split}
  \label{V_rxy:b}
\end{gather}
To find the $t$-derivatives we differentiate the orthonormality relation
\begin{equation}
  0 = -{\dot{w_0} \over w_0}\delta_{mn} 
    + \int {dz \over 2\pi iz} \tilde{w}[\dot{\phi}_m+z\phi_m]\overline{\phi}_n
    + \int {dz \over 2\pi iz} \tilde{w}\phi_m\overline{\dot{\phi}}_n
\end{equation}
and the case $ n \leq m-2 $ indicates that $ \dot{\phi_m}+z\phi_m $ has no
components in $ \phi_n $ so
\begin{equation}
  \dot{\phi}_m+z\phi_m = \bar{a}_m\phi_{m+1}+\bar{d}_m\phi_m+\bar{e}_m\phi_{m-1}.
\end{equation} 
Consideration of the coefficients of the highest power in $ z $ gives 
$ \bar{a}_m = \kappa_m/\kappa_{m+1} $, and the case $ n=m-1 $ shows $ \bar{e}_m=0 $.
The $t$-derivative of the orthogonal polynomial is
\begin{equation}
  \dot{\phi}_n 
  = \left[\bar{d}_n+{\kappa_n\phi_{n+1}(0) \over \kappa_{n+1}\phi_{n}(0)}\right]\phi_n
    - {\kappa_{n-1}\phi_{n+1}(0) \over \kappa_{n+1}\phi_{n}(0)}z\phi_{n-1} 
\end{equation} 
and this implies
\begin{equation}
   {\dot{r}_n \over r_n} = r_{n+1}\left({1 \over r_n}-\bar{r}_n\right).
\label{V_rdot:a}
\end{equation}
Next we examine the case $ n \geq m+2 $ and find
\begin{equation}
  \dot{\bar{\phi}}_n = b_n\bar{\phi}_{n}+c_n\bar{\phi}_{n-1}.
\end{equation} 
Taking the case $ n=m+1 $ we infer $ c_n=-\kappa_{n-1}/\kappa_{n} $. The 
$t$-derivative of the orthogonal polynomial can then be written as
\begin{equation}
  \dot{\bar{\phi}}_n = b_n\bar{\phi}_n - {\kappa_{n-1} \over \kappa_{n}}\bar{\phi}_{n-1}, 
\end{equation}
from which we deduce
\begin{equation}
   {\dot{\bar{r}}_n \over \bar{r}_n} 
   = -\bar{r}_{n-1}\left({1 \over \bar{r}_n}-r_n\right).
\label{V_rdot:b}
\end{equation}
Substituting (\ref{V_rdot:a}), (\ref{V_rdot:b}) in (\ref{V_rxy:a}), (\ref{V_rxy:b})
gives (\ref{VH_Ops:b},\ref{VH_Ops:c}) respectively.

To derive (\ref{V_xfm:a}), multiply both sides of (\ref{VH_Ops:b}) by 
$ (1-y_N)/y_N $, then substitute for 
\begin{equation*}
  X{1-y_N \over y_N}, \qquad X := x_N-{t \over y_N}-\nu-\half\mu,
\end{equation*}
using (\ref{VH_Ops:a}) to deduce 
\begin{equation*}
  \left[ (N+\nu)(1-r_N\bar{r}_N)-N \right]
  \left( 1-{1 \over y_Nr_N\bar{r}_N} \right)-N{1-y_N \over y_N} 
  = t{r_{N+1} \over r_{N}}(1-r_N\bar{r}_N){1-y_N \over y_N} .
\end{equation*}
Solving this equation for $ y_N $ gives (\ref{V_xfm:a}). For the derivation of 
(\ref{V_xfm:b}) we write (\ref{VH_Ops:a}) in the form
\begin{equation*}
  (X+\nu)(1-y_N) = -(N+\nu)r_N\bar{r}_Ny_N+\nu .
\end{equation*}
Substituting (\ref{V_xfm:a}) for $ y_N $ and simplifying we obtain (\ref{V_xfm:b}). 
\end{proof}

The Hamiltonian variables $ q_N, p_N $ or $ x_N, y_N $ in the \PV theory go over
to those in the \PIIIa theory under the replacements (\ref{V_to_III}) and upon 
taking the limit $ \nu \to \infty $
\begin{gather} 
   q^{\rm V}_N 
  \mathop{\rightarrow}\limits_{\nu \to \infty} 1-p^{\rm III'}_N ,
  \qquad
   p^{\rm V}_N
  \mathop{\rightarrow}\limits_{\nu \to \infty} q^{\rm III'}_N
  \\
   x^{\rm V}_N
  \mathop{\rightarrow}\limits_{\nu \to \infty}
    \half\mu+q^{\rm III'}_N(1-p^{\rm III'}_N) ,
  \qquad
   y^{\rm V}_N
  \mathop{\rightarrow}\limits_{\nu \to \infty} {1\over 1-p^{\rm III'}_N} .
\end{gather} 
Using these transitions we find that the recurrence relations (\ref{V_hRecur:a}),
(\ref{V_hRecur:b}) reduce to (\ref{III_hRecur:a}) upon using (\ref{III_hRecur:c})
(which follows from (\ref{III_hRecur:b}) and (\ref{III_hRecur:a})), 
and to (\ref{III_hRecur:a}) respectively.

In addition to the formula (\ref{V_xfm:a}) for $ y_N $, we can obtain a different
formula by making use of (\ref{VH_Ops:c}). This allows a pair of $ 1/1 $ order 
difference equations for $ r_N, \bar{r}_N $ to be deduced, thus reducing the 
$ 2/1 $ system of Proposition \ref{V_rProp} down to the same order as the coupled 
system (\ref{V_hRecur:a}), (\ref{V_hRecur:b}) satisfied by $ x_N, y_N $.

\begin{theorem}
The reflection coefficients satisfy the coupled $ 1/1 $ order recurrence relations 
\begin{gather}
    (1-r_N\bar{r}_N)[tr_{N+1}+(N+\nu)r_N][t\bar{r}_{N-1}+(N+\nu)\bar{r}_N]
   \label{V_rRecur:d} \\
   = [(N+\nu)r_N\bar{r}_N+\mu][\nu-(N+\nu)r_N\bar{r}_N]
   \nonumber \\
     t^2r^2_{N}\bar{r}^2_{N-1}+t(\nu-\mu-t)r_{N}\bar{r}_{N-1}
     -(N+\nu)(N-1+\nu)\bar{r}_{N}r_{N-1}
   \label{V_rRecur:e} \\
     -(N-1+\nu)tr_{N-1}\bar{r}_{N-1}-(N+\nu)tr_{N}\bar{r}_{N} -\mu\nu = 0
   \nonumber
\end{gather}
\end{theorem}
\begin{proof}
Multiplying both sides of (\ref{VH_Ops:c}) by $ 1/y_N $ and substituting for 
$ X(1-y_N)/y_N $ using (\ref{VH_Ops:a}) then solving for $ y_N $ shows
\begin{equation}
  y_N = {N+\mu+\nu+t{\displaystyle\bar{r}_{N-1} 
                     \over \displaystyle\overset{}{\bar{r}_{N}}}
                    (1-r_N\bar{r}_N)
      \over \mu+(N+\nu)r_N\bar{r}_N} . 
\end{equation}
Equating this with (\ref{V_xfm:a}) and solving for $ r_{N+1} $ gives 
(\ref{V_rRecur:d}). The second recurrence follows by eliminating $ r_{N+1} $ between 
(\ref{V_rRecur:d}) and (\ref{V_rRecur:a}).
\end{proof}

As with the other recurrences for $ r_N, \bar{r}_N $ in the \PV system, we find
(\ref{V_rRecur:d}), (\ref{V_rRecur:e}) under the replacements (\ref{V_to_III})
and taking the limit $ \nu \to \infty $ assume the forms of the recurrences in the 
\PIIIa system (\ref{III_rRecur:a}), (\ref{III_rRecur:c}) respectively.

\section{Concluding Remarks}
\setcounter{equation}{0}

In our work \cite{FW_2003a}, in addition to the $ U(N) $ averages (\ref{III_Toep}),
(\ref{V_Toep}), an $ N$-recurrence was also obtained for 
\begin{equation} 
   \tau^{\rm VI}[N](t;\mu,\omega_1,\omega_2;\xi) 
  = \Big\langle \prod^N_{l=1}(1-\xi\chi^{(l)}_{(\pi-\phi,\pi)})
                 e^{\omega_2\theta_l}|1+z_l|^{2\omega_1}
                 \left({1\over tz_l}\right)^{\mu}(1+tz_l)^{2\mu}
    \Big\rangle_{U(N)} ,
\label{VI_Toep}
\end{equation} 
where $ \chi^{(l)}_{J} = 1 $ for  $ \theta_l \in J $, $ \chi^{(l)}_{J} = 0 $ 
otherwise. As the notation suggests, this is a known $ \tau$-function for a \PVI
system \cite{FW_2002b}. Both $ \tau^{\rm III'}[N] $ and $ \tau^{\rm V}[N] $ can be
obtained as degenerations of (\ref{VI_Toep}), or equivalently the weights 
(\ref{SC_wgt}) can be obtained as limiting cases of the "master" semi-classical
weight function underlying (\ref{VI_Toep}).
For general parameters, a recurrence scheme based on the
discrete Painlev\'e equation associated with the degeneration of the rational 
surface $ D^{(1)}_4 \to D^{(1)}_5 $ (discrete Painlev\'e {\rm V}), was given 
as a consequence of the Painlev\'e system theory of \PVI (recurrence schemes for
special cases of the parameters have also been given in \cite{Bo_2001},
\cite{BB_2002}). It is also true that an $ N$-recurrence for (\ref{VI_Toep}) can be
deduced from the Toeplitz lattice approach of Adler and van Moerbeke 
\cite{AvM_2002}. But here we have found that in the cases of the $ U(N) $ averages
(\ref{III_Toep}),(\ref{V_Toep}) the approach of \cite{AvM_2002} leads to equivalent
results as do those obtained from an orthogonal polynomial approach. One therefore
suspects the same will be true in relation to (\ref{VI_Toep}), and that furthermore
the corresponding recurrences are transformed versions of the discrete Painlev\'e 
{\rm V} equation found in \cite{FW_2002b}. This is indeed the case, but the 
details do not fit well with the scheme of the present paper (in particular the 
Freud approach to the recurrences for $ r_n, \bar{r}_n $ is now inadequate) so 
will be reported elsewhere.

Another point of interest relates to the $N$-recurrences for Hermitian matrix
(as opposed to unitary matrix) averages in which the weight function is a 
$ q$-generalisation of a classical weight function. In \cite{BB_2002} it is shown
that the method of Borodin leads to $q$-discrete Painlev\'e equations. Can one
obtain the $q$-discrete Painlev\'e equations from an orthogonal polynomial
approach? 

\subsection*{Acknowledgment}
This research has been supported by the Australian Research Council.

\bibliographystyle{amsplain}
\bibliography{moment,nonlinear,random_matrices}

%\end{spacing}
\end{document}